\documentclass[12pt]{article}
\usepackage{axodraw,bbold}

\catcode`@=12
\begin{document}
\thispagestyle{empty}
\begin{flushright} 
UCRHEP-T422\\ 
October 2006\
\end{flushright}
\vspace{0.5in}
\begin{center}
{\LARGE	\bf Supersymmetric $A_4 \times Z_3$ and $A_4$ Realizations\\ of 
Neutrino Tribimaximal Mixing\\ Without and With Corrections\\}
\vspace{1.5in}
{\bf Ernest Ma\\}
\vspace{0.2in}
{\sl Physics Department, University of California, Riverside, 
California 92521\\}
\vspace{1.5in}
\end{center}

\begin{abstract}\
In an improved application of the tetrahedral symmetry $A_4$ first introduced 
by Ma and Rajasekaran, supplemented by the discrete symmetry $Z_3$ as well as 
supersymmetry, a two-parameter form of the neutrino mass matrix is derived 
which exhibits the tribimaximal mixing of Harrison, Perkins, and Scott.  
This form is the same one obtained previously by Altarelli and Feruglio, 
and the inverse of that obtained by Babu and He.  If only $A_4$ is used, 
then corrections appear, making $\tan^2 \theta_{12}$ different from 0.5, 
without changing significantly $\sin^2 2 \theta_{23}$ from one or 
$\theta_{13}$ from zero.
\end{abstract}

\newpage
\baselineskip 24pt

Based on present neutrino-oscillation data, the neutrino mixing matrix 
$U_{\alpha i}$ linking the charged leptons ($\alpha = e, \mu, \tau$) to 
the neutrino mass eigenstates ($i = 1, 2, 3$) is determined to a large 
extent \cite{fogli06}.  In particular, a good approximate description is 
that of the so-called tribimaximal mixing of Harrison, Perkins, and Scott 
\cite{hps02}, i.e.
\begin{equation}
U_{\alpha i} = \pmatrix{ \sqrt{2/3} & \sqrt{1/3} & 0 \cr -\sqrt{1/6} & 
\sqrt{1/3} & -\sqrt{1/2} \cr -\sqrt{1/6} & \sqrt{1/3} & \sqrt{1/2}}.
\end{equation}
Since $U_{\alpha i}$ is the mismatch of the two unitary matrices 
diagonalizing ${\cal M}_l {\cal M}_l^\dagger$ and ${\cal M}_\nu 
{\cal M}_\nu^\dagger$, it is interesting to note that Eq.~(1) can be 
rewritten as
\begin{equation}
U_{\alpha i} = U_l^\dagger \pmatrix{1 & 0 & 0 \cr 0 & 1/\sqrt 2 & -1/\sqrt 2 
\cr 0 & 1/\sqrt 2 & 1/\sqrt 2} \pmatrix{0 & 1 & 0 \cr 1 & 0 & 0 \cr 0 & 0 & i},
\end{equation}
where
\begin{equation}
U_l = {1 \over \sqrt 3} \pmatrix{1 & 1 & 1 \cr 1 & \omega & \omega^2 \cr 
1 & \omega^2 & \omega},
\end{equation}
with $\omega = \exp(2 \pi i/3) = -1/2 + i \sqrt{3}/2$.  This $U_l$ was first 
derived \cite{mr01,bmv03} in the context of the discrete family symmetry 
$A_4$, although remarkably enough, it was already conjectured by 
Cabibbo \cite{c78} and Wolfenstein \cite{w78} before any experimental data 
were available to be $U_{\alpha i}$ itself many years ago.  In the context of 
Eq.~(2), this would have been the case if ${\cal M}_\nu$ were diagonal, but 
to obtain Eq.~(1), ${\cal M}_\nu$ should instead be of the form \cite{m04}
\begin{equation}
{\cal M}_\nu = \pmatrix{a+2b & 0 & 0 \cr 0 & a-b & d \cr 0 & d & a-b},
\end{equation}
with eigenvalues $m_1 = a-b+d$, $m_2 = a+2b$, and $m_3 = -a+b+d$. 
The exact form of this mixing matrix has been derived recently in a 
number of papers, using $A_4$ \cite{af05,bh05,af06,m06_1} as well as $B_4$ 
\cite{gl06}.  In particular, the special cases of $b=0$ \cite{af05,af06}, 
and $d^2=3b(b-a)$ \cite{bh05} have been obtained.  Here a new specific 
supersymmetric $A_4 \times Z_3$ model is proposed, resulting in Eq.~(4) 
with $b=0$ as well.  It differs from all previous such models in that 
the $A_4$ assignments of the charged-lepton sector are not those of  
the original proposal \cite{mr01,bmv03}, but rather those of Ref.~\cite{m06_2}.

The non-Abelian finite group $A_4$ is the symmetry group of the even 
permutation of four objects.  It is also the symmetry group of the regular 
tetrahedron, one of five perfect geometric solids which was identified by 
Plato with the Greek element ``fire'' \cite{plato}.  There are twelve group 
elements and four irreducible representations: \underline{1}, 
\underline{1}$'$, \underline{1}$''$, and \underline{3}.  Let $a_{1,2,3}$ 
and $b_{1,2,3}$ transform as \underline{3} under $A_4$, then \cite{fuji}
\begin{eqnarray}
&& a_1 b_1 + a_2 b_2 + a_3 b_3 \sim \underline{1}, \\ 
&& a_1 b_1 + \omega^2 a_2 b_2 + \omega a_3 b_3 \sim \underline{1}', \\ 
&& a_1 b_1 + \omega a_2 b_2 + \omega^2 a_3 b_3 \sim \underline{1}'', \\ 
&& (a_2 b_3, a_3 b_1, a_1 b_2) \sim \underline{3}, \\ 
&& (a_3 b_2, a_1 b_3, a_2 b_1) \sim \underline{3}, 
\end{eqnarray}
The superfield content of this model is given in Table. 1.

\begin{table}[htb]
\caption{Particle content of proposed model.}
\begin{center}
\begin{tabular}{|c|c|c|c|}
\hline 
Superfield & $SU(2) \times U(1)$ & $A_4$ & $Z_3$ \\ 
\hline
$L_i = (\nu_i,l_i)$ & $(2,-1/2)$ & \underline{3} & 1 \\ 
$l^c_i$ & $(1,1)$ & \underline{3} & 1 \\ 
$\Phi_1 = (\phi^0_1,\phi^-_1)$ & $(2,-1/2)$ & \underline{1} & 1 \\ 
$\Phi_2 = (\phi^+_2,\phi^0_2)$ & $(2,1/2)$ & \underline{1} & 1 \\ 
\hline
$\sigma_i$ & $(1,0)$ & \underline{3} & 1 \\ 
$\chi_i$ & $(1,0)$ & \underline{3} & $\omega$ \\ 
$\zeta$ & $(1,0)$ & \underline{1} & $\omega$ \\ 
$\xi_1 = (\xi^{++}_1,\xi^+_1,\xi^0_1)$ & $(3,1)$ & \underline{1} & 
$\omega^2$ \\ 
$\xi_2 = (\xi^0_2,\xi^-_2,\xi^{--}_2)$ & $(3,-1)$ & \underline{1} & 
$\omega^2$ \\ 
\hline
\end{tabular}
\end{center}
\end{table}

In the charged-lepton sector, the only allowed trilinear Yukawa term is 
$L_i l^c_i \Phi_1$.  However, using the singlets $\sigma_i$, the effective 
quadrilinear term $L_i l^c_j \sigma_k \Phi_1$ may also be written, resulting 
in
\begin{equation}
{\cal M}_l = \pmatrix{h_0 v_0 & h_1 v_3 & h_2 v_2 \cr h_2 v_3 & h_0 v_0 & 
h_1 v_1 \cr h_1 v_2 & h_2 v_1 & h_2 v_0},
\end{equation}
where $v_0 = \langle \phi_1^0 \rangle$ and $v_i = \langle \sigma_i \rangle 
\langle \phi_1^0 \rangle / \Lambda$, with $\Lambda$ being a very high scale.  
As shown recently \cite{m06_2}, for $v_1=v_2=v_3=v$, this mass matrix is 
exactly diagonalized by $U_l$ of Eq.~(3):
\begin{equation}
{\cal M}_l = U_l \pmatrix{h_0 v_0 + (h_1+h_2)v & 0 & 0 \cr 0 & h_0 v_0 + 
(\omega h_1 + \omega^2 h_2)v & 0 \cr 0 & 0 & h_0 v_0 + (\omega^2 h_1 + 
\omega h_2)v} U_l^\dagger.
\end{equation}
This establishes the first step in obtaining Eq.~(1).

In the neutrino sector, the allowed terms are $L_i L_i \zeta \xi_1$ and 
$L_i L_j \chi_k \xi_1$.  The form of ${\cal M}_\nu$ is then the same 
as ${\cal M}_l$ except that it must be symmetric, i.e.
\begin{equation}
{\cal M}_\nu = \pmatrix{f_0 u_0 & f_1 u_3 & f_1 u_2 \cr f_1 u_3 & f_0 u_0 & 
f_1 u_1 \cr f_1 u_2 & f_1 u_1 & f_0 u_0},
\end{equation}
where $u_0 = \langle \zeta \rangle \langle \xi_1^0 \rangle / \Lambda$ and 
$u_i = \langle \chi_i \rangle \langle \xi_1^0 \rangle / \Lambda$.  For 
$u_2 = u_3 = 0$, this reduces to Eq.~(4) with $b=0$, and Eq.~(1) is 
obtained \cite{m04}.

The problem boils down now to showing that $A_4$ may be broken spontaneously 
at the same time in two different directions: (1,1,1), i.e. $v_1=v_2=v_3$, 
and (1,0,0), i.e. $u_2=u_3=0$.  This must be accomplished using the 
superpotential of the singlet superfields $\sigma_i$, $\zeta$, and $\chi_i$ 
without breaking the supersymmetry \cite{m06_3}.  The most general such 
superpotential allowing the soft breaking of $Z_3$ but not $A_4$ is given by 
\begin{eqnarray}
W &=& {1 \over 2} m_\sigma (\sigma_1^2 + \sigma_2^2 + 
\sigma_3^2) + {1 \over 2} m_\zeta \zeta^2 + {1 \over 2} m_\chi 
(\chi_1^2 + \chi_2^2 + \chi_3^2) \nonumber \\ 
&+& \lambda_1 \sigma_1 \sigma_2 \sigma_3 + \lambda_2 \zeta^3 + 
\lambda_3 \zeta (\chi_1^2 + \chi_2^2 + \chi_3^2) + \lambda_4 \chi_1 \chi_2 
\chi_3.
\end{eqnarray}
To preserve the supersymmetry of the complete theory at this high scale, 
a solution must exist for which the minimum of the resulting scalar 
potential, i.e.
\begin{eqnarray}
V &=& |m_\sigma \sigma_1 + \lambda_1 \sigma_2 \sigma_3|^2 + 
|m_\sigma \sigma_2 + \lambda_1 \sigma_1 \sigma_3|^2 + 
|m_\sigma \sigma_3 + \lambda_1 \sigma_1 \sigma_2|^2 \nonumber \\ 
&+& |m_\zeta \zeta + 3 \lambda_2 \zeta^2 + \lambda_3 (\chi_1^2 + \chi_2^2 + 
\chi_3^2)|^2 + 
|m_\chi \chi_1 + 2 \lambda_3 \zeta \chi_1 + \lambda_4 \chi_2 \chi_3|^2 
\nonumber \\ 
&+& |m_\chi \chi_2 + 2 \lambda_3 \zeta \chi_2 + \lambda_4 \chi_1 \chi_3|^2 + 
|m_\chi \chi_3 + 2 \lambda_3 \zeta \chi_3 + \lambda_4 \chi_1 \chi_2|^2,
\end{eqnarray}
is zero.  It is easy to check that
\begin{eqnarray}
&& \langle \sigma_1 \rangle = \langle \sigma_2 \rangle = \langle \sigma_3 
\rangle = \langle \sigma \rangle = {-m_\sigma \over \lambda_1}, ~~~ 
\langle \zeta \rangle = {-m_\chi \over 2 \lambda_3}, \\ 
&& \langle \chi_1 \rangle^2 = {m_\chi (2 \lambda_3 m_\zeta - 3 \lambda_2 
m_\chi) \over 4 \lambda_3^2}, ~~~ \langle \chi_2 \rangle = 
\langle \chi_3 \rangle = 0,
\end{eqnarray}
is such a solution.

Below this $A_4 \times Z_3$ breaking scale, the superpotential includes 
the term
\begin{eqnarray}
\Phi_1 \{ h_0 (L_1 l^c_1 + L_2 l^c_2 + L_3 l^c_3) 
 + h'_1 (L_1 l^c_2 + L_2 l^c_3 + L_3 l^c_1)  
 + h'_2 (L_2 l^c_1 + L_3 l^c_2 + L_1 l^c_3) \},
\end{eqnarray}
where $h'_1 = h_1 \langle \sigma \rangle / \Lambda$, $h'_2 = h_2 \langle 
\sigma \rangle / \Lambda$, as well as
\begin{equation}
\xi_1 \left\{ {1 \over 2} f'_0 (L_1 L_1 + L_2 L_2 + L_3 L_3) + f'_1 L_2 L_3 
\right\},
\end{equation}
where $f'_0 = f_0 \langle \zeta \rangle / \Lambda$, $f'_1 = f_1 \langle 
\chi_1 \rangle / \Lambda$.  The quadrilinear terms $\zeta \xi_1 \Phi_1 \Phi_1 
/ \Lambda$ and $\zeta \xi_2 \Phi_2 \Phi_2 / \Lambda$ become trilinear terms 
$\xi_1 \Phi_1 \Phi_1$ and $\xi_2 \Phi_2 \Phi_2$, allowing $\xi_{1,2}$ to 
acquire small vacuum expectation values \cite{ms98,hms01}.

This is thus another version of a successful derivation of tribimaximal 
mixing using $A_4$ \cite{others}, but the predicted value of $\tan^2 
\theta_{12} = 0.5$ is not the central value of present experimental data: 
$\tan^2 \theta_{12} = 0.45 \pm 0.05$.  To obtain a deviation from $\tan^2 
\theta_{12} = 0.5$ and still retain some kind of symmetry, one is tempted 
to fix $\nu_2 = (\nu_e + \nu_\mu + \nu_\tau)/\sqrt 3$ and let $\nu_1$ mix 
with $\nu_3$. However, as shown already in Ref.~\cite{m04}, this would only 
make $\tan^2 \theta_{12} > 0.5$.  A better idea is required.

Another complication with tribimaximal mixing in the context of $A_4$ is 
the necessity of auxiliary symmetries such as the $Z_3$ discussed here. 
Consider then the case of $A_4$ by itself.  The most general superpotential 
consisting of $\sigma_i$ and $\chi_i$, both transforming as triplets under 
$A_4$, is then given by
\begin{eqnarray}
W &=& {1 \over 2} m_1 (\sigma_1^2 + \sigma_2^2 + \sigma_3^2) 
+ {1 \over 2} m_2 (\chi_1^2 + \chi_2^2 + \chi_3^2) 
+ m_3 (\sigma_1 \chi_1 + \sigma_2 \chi_2 + \sigma_3 \chi_3) 
\nonumber \\ 
&+& \lambda_1 \sigma_1 \sigma_2 \sigma_3 + \lambda_2 (\sigma_1 \chi_2 \chi_3 
+ \sigma_2 \chi_3 \chi_1 + \sigma_3 \chi_1 \chi_2) 
\nonumber \\
&+& \lambda_3 (\chi_1 \sigma_2 \sigma_3 + \chi_2 \sigma_3 \sigma_1 
+ \chi_3 \sigma_1 \sigma_2) + \lambda_4 \chi_1 \chi_2 \chi_3.
\end{eqnarray}
It is straightforward to show that
\begin{equation}
\langle \chi_2 \rangle = \langle \chi_3 \rangle = 0, ~~~ 
\langle \sigma_2 \rangle = \langle \sigma_3 \rangle \neq 0,
\end{equation}
is a solution which breaks $A_4$ but not the supersymmetry, provided that
\begin{equation}
\lambda_2 m_1 + \lambda_1 m_2 = 2 \lambda_3 m_3.
\end{equation}
In that case,
\begin{equation}
\langle \chi_1 \rangle = {\lambda_1 m_3 - \lambda_3 m_1 \over \lambda_3^2 - 
\lambda_1 \lambda_2}, ~~~
\langle \sigma_1 \rangle = {\lambda_2 m_1 - \lambda_3 m_3 \over \lambda_3^2 - 
\lambda_1 \lambda_2}, ~~~
\langle \sigma_{2,3} \rangle^2 = {m_1 m_2 - m_3^2 \over \lambda_3^2 - 
\lambda_1 \lambda_2}.
\end{equation}
Note that $\langle \sigma_1 \rangle = \langle \sigma_{2,3} \rangle$ is not 
allowed, because that would imply $\lambda_1 m_3 = \lambda_3 m_1$, i.e. 
$\langle \chi_1 \rangle = 0$ necessarily.  As for the condition of Eq.~(21), 
although it is {\it ad hoc}, since it allows the supersymmetry to be exact 
even after the breaking of $A_4$, it will be protected against radiative 
corrections until the supersymmetry itself is broken.

Since $\langle \sigma_1 \rangle \neq \langle \sigma_{2,3} \rangle$, the 
charged-lepton mass matrix is now modified.  Using the original $A_4$ 
assignment of \underline{1}, \underline{1}$'$, \underline{1}$''$ for 
$l^c_i$, it is then of the form
\begin{equation}
{\cal M}_l = \pmatrix{h_e v_1 & h_\mu v_1 & h_\tau v_1 \cr h_e v_2 & 
h_\mu \omega v_2 & h_\tau \omega^2 v_2 \cr h_e v_2 & h_\mu \omega^2 v_2 & 
h_\tau \omega v_2}.
\end{equation}
Applying the $U_l$ of Eq.~(3) and using the phenomenological hierarchy 
$h_e << h_\mu << h_\tau$, it is easily shown, to first approximation, that 
the tribimaximal $U_{\alpha i}$ of Eq.~(1) is multiplied on the left by
\begin{equation}
R = \pmatrix{1 & -r & -r \cr r & 1 & -r \cr r & r & 1},
\end{equation}
where $r = (v_1-v_2)/(v_1+2v_2)$.  Hence the corrected mixing matrix is 
given by
\begin{equation}
U_{\alpha i} = \pmatrix{ \sqrt{2/3}(1+r) & \sqrt{1/3}(1-2r) & 0 \cr 
-\sqrt{1/6}(1-3r) & \sqrt{1/3} & -\sqrt{1/2}(1+r) \cr -\sqrt{1/6}(1-r) & 
\sqrt{1/3}(1+2r) & \sqrt{1/2}(1-r)}.
\end{equation}
Therefore,
\begin{equation}
\tan^2 \theta_{12} \simeq {1 \over 2} - 3r, ~~~ 
\tan^2 \theta_{23} \simeq 1+4r, ~~~ \theta_{13} \simeq 0.
\end{equation}
For example, let $r = 0.02$, then $\tan^2 \theta_{12} \simeq 0.44$, whereas 
$\tan^2 \theta_{23} \simeq 1.08$ which is equivalent to $\sin^2 2 \theta_{23} 
\simeq 0.9985$.  A better match to the data is thus obtained.

In conclusion, it has been shown in this paper that neutrino tribimaximal 
mixing is possible in a simple supersymmetric model based on $A_4 \times Z_3$, 
but with $A_4$ alone, corrections will occur, such that $\tan^2 \theta_{12}$ 
may deviate significantly from 0.5 without affecting much the predictions 
$\sin^2 2 \theta_{23} = 1$ and $\theta_{13} = 0$.

This work was supported in part by the U.~S.~Department of Energy under Grant 
No.~DE-FG03-94ER40837.

\bibliographystyle{unsrt}

\end{document}